\documentstyle[aps]{revtex}
\begin{document}
\title{Some Remarks on the Fragmentation of Bose Condensates}
\author{R. W. Spekkens and J. E. Sipe}
\address{Department of Physics, University of Toronto, Toronto,\\
Ontario, Canada M5S 1A7}
\date{Sept. 15, 1997}
\maketitle

\begin{abstract}
We show how fragmentation of a Bose-Einstein condensate can occur given
repulsive inter-particle interactions and a non-uniform external potential.
\end{abstract}

\section{Introduction}

It is customary to approximate the $N$-body ground state of a system of
spinless bosons by the Hartree-Fock state 
\[
\left| N,\phi _0\right\rangle =\frac{\left( \int d^3{\bf r}\phi _0({\bf r})%
\hat \Psi ^{\dag }({\bf r})\right) ^N}{\sqrt{N!}}\left| \text{vac}%
\right\rangle , 
\]
where $\phi _0({\bf r})$ is a normalized single particle wavefunction \cite
{coherent state}. This describes an accumulation of $N$ particles in the
wavefunction $\phi _0({\bf r}).$ One can, however, also consider states of
the form 
\begin{equation}
\left| N_1,\phi _1;N_2,\phi _2\right\rangle =\frac{\left( \int d^3{\bf r}%
\phi _1({\bf r})\hat \Psi ^{\dag }({\bf r})\right) ^{N_1}}{\sqrt{N_1!}}\frac{%
\left( \int d^3{\bf r}\phi _2({\bf r})\hat \Psi ^{\dag }({\bf r})\right)
^{N_2}}{\sqrt{N_2!}}\left| \text{vac}\right\rangle ,
\end{equation}
where $\phi _1({\bf r})$ and $\phi _2({\bf r})$ are normalized and
orthogonal single-particle wavefunctions, and where $N_1+N_2=N$. Such a
state describes an accumulation of $N_1$ particles in the wavefunction $\phi
_1({\bf r}),$ and $N_2$ particles in $\phi _2({\bf r}).$ We distinguish
states of the form $\left| N,\phi _0\right\rangle $ from those of the form $%
\left| N_1,\phi _1;N_2,\phi _2\right\rangle $ by designating the first as
`single condensates' and the second as `dual condensates'. It is evident
that one can also introduce states describing the accumulation of particles
in an arbitrary number of orthogonal single-particle wavefunctions. Such
states can be referred to collectively as `fragmented condensates' \cite
{Nozieres}. In this paper, we show how a non-uniform trap potential can
encourage fragmentation. In particular, we demonstrate that in a double
minimum potential a dual condensate can have an expectation value of the
energy smaller than that of a single condensate, and therefore can
constitute a better approximation to the fully interacting ground state and
a better starting point for treatments beyond Hartree-Fock.

This result may be surprising to some and obvious to others. Specifically,
it seems remarkable in light of Nozi\`{e}res' argument against fragmentation
of condensates \cite{Nozieres}. On the other hand, it seems natural to
describe bosons in well-separated traps by a product of single condensate
states for each trap; for two traps, such a product state would be a dual
condensate wherein each of $\phi _{1}({\bf r})$ and $\phi _{2}({\bf r}) $ is
centered in only one of the traps. Indeed, this has been assumed implicitly
in the literature when discussing the interference of condensates \cite
{Castin and Dalibard}. Our analysis justifies this intuition and specifies
the circumstances under which Nozi\`{e}res' argument is inapplicable.

\section{Desiderata for fragmentation}

We consider a two-particle interaction that is local and completely
repulsive: $V({\bf r},{\bf r}^{\prime })=g\delta ({\bf r}-{\bf r}^{\prime
}), $ The interaction strength is given by $g=4\pi a_{sc}\hbar ^{2}/m,$
where $a_{sc}$ is the s-wave scattering length, assumed to be positive, and
where $m $ is the mass of the particle. For simplicity we restrict ourselves
to wavefunctions $\phi _{0}({\bf r}),$ $\phi _{1}({\bf r}),$ and $\phi _{2}(%
{\bf r})$ that are real. Consider the expectation value of the energy for a
single condensate $\left| N,\phi _{0}\right\rangle ,$%
\begin{equation}
E_{\text{s}}=N\epsilon (\phi _{0})+\frac{1}{2}gN(N-1)\int \phi _{0}^{4}({\bf %
r})d^{3}{\bf r,}  \label{H sl}
\end{equation}
where $\epsilon (\phi )=\int \phi ({\bf r})\left[ -\frac{\hbar ^{2}}{2m}{\bf %
\nabla }^{2}+U({\bf r})\right] \phi ({\bf r})d^{3}{\bf r}$ is the single
particle energy associated with $\phi ({\bf r}),$ and $U({\bf r)}$ is the
trap potential. For the dual condensate $\left| N_{1,}\phi _{1};N_{2},\phi
_{2}\right\rangle ,$ 
\begin{eqnarray}
E_{\text{d}} &=&N_{1}\epsilon (\phi _{1})+N_{2}\epsilon (\phi _{2})+\frac{1}{%
2}gN_{1}(N_{1}-1)\int \phi _{1}^{4}({\bf r})d^{3}r  \nonumber \\
&&+\frac{1}{2}gN_{2}(N_{2}-1)\int \phi _{2}^{4}({\bf r})d^{3}r+2gN_{1}N_{2}%
\int \phi _{1}^{2}({\bf r})\phi _{2}^{2}({\bf r})d^{3}r.  \label{H dl}
\end{eqnarray}

If we assume that the particle densities are similar throughout the single
and dual condensates, $N_{1}\phi _{1}^{2}({\bf r})+N_{2}\phi _{2}^{2}({\bf r}%
)\simeq N\phi _{0}^{2}({\bf r}),$ then the pieces of the interaction
energies that scale as $N^{2}$ are 
\begin{eqnarray}
E_{\text{s}}^{\text{quad}} &=&\frac{1}{2}gN^{2}\int \phi _{0}^{4}({\bf r}%
)d^{3}{\bf r}  \nonumber \\
E_{\text{d}}^{\text{quad}} &=&\frac{1}{2}gN_{1}^{2}\int \phi _{1}^{4}({\bf r}%
)d^{3}{\bf r+}\frac{1}{2}gN_{2}^{2}\int \phi _{2}^{4}({\bf r})d^{3}{\bf r+2}%
gN_{1}N_{2}\int \phi _{1}^{2}({\bf r})\phi _{2}^{2}({\bf r})d^{3}r  \nonumber
\\
&\simeq &\frac{1}{2}gN^{2}\int \phi _{0}^{4}({\bf r})d^{3}{\bf r+}%
gN_{1}N_{2}\int \phi _{1}^{2}({\bf r})\phi _{2}^{2}({\bf r})d^{3}r
\label{int energy}
\end{eqnarray}
We can think of this in the following way:{\it \ }while the Hartree
contributions to the interaction energy of the single and dual condensates
are essentially equal, the interaction energy of the dual condensate also
includes an exchange term, proportional to $N_{1}N_{2},$ arising from
symmetrization.

This extra exchange energy was identified by Nozi\`{e}res \cite{Nozieres}
and is the basis of his argument against fragmentation of the condensate,
since in the limit of a macroscopic number of particles, $E_{\text{d}}^{%
\text{quad}}$ exceeds $E_{\text{s}}^{\text{quad}}$ by a macroscopic energy.
However, this argument assumes that the wavefunctions have a similar spatial
extent $\phi _{1}^{2}({\bf r})\simeq \phi _{2}^{2}({\bf r})\simeq \phi
_{0}^{2}({\bf r}).$ If instead one considers a dual condensate wherein $\phi
_{1}$ and $\phi _{2}$ have little density overlap, then the exchange term
may be small compared to other terms in (\ref{int energy}) that are linear
in $N,$ and we must consider the relative weights of all the terms.

The energies $E_{\text{d}}^{\text{quad}}$ and $E_{\text{s}}^{\text{quad}}$
overestimate the interaction energy since they incorporate a particle's
interaction with itself. Thus, to obtain the proper interaction energy, we
must add to $E_{\text{s}}^{\text{quad}}$ a term linear in $N$ 
\begin{equation}
E_{\text{s}}^{\text{lin}}=-\frac{1}{2}gN\int \phi _{0}^{4}({\bf r})d^{3}{\bf %
r.}
\end{equation}
Similarly, we must add to $E_{\text{d}}^{\text{quad}}$ the linear term 
\begin{equation}
E_{\text{d}}^{\text{lin}}=-\frac{1}{2}gN_{1}\int \phi _{1}^{4}({\bf r})d^{3}%
{\bf r}-\frac{1}{2}gN_{2}\int \phi _{2}^{4}({\bf r})d^{3}{\bf r.}
\end{equation}
If $\phi _{1}$ and $\phi _{2}$ are more localized than $\phi _{0},$ then $%
\int \phi _{1,2}^{4}({\bf r})d^{3}r>\int \phi _{0}^{4}({\bf r})d^{3}{\bf r.}$
It follows that $E_{\text{d}}^{\text{lin}}-E_{\text{s}}^{\text{lin}}<$ $-%
\frac{1}{2}gN\int \phi _{0}^{4}({\bf r})d^{3}{\bf r.}$ Thus, the linear
piece of the interaction energy favors the dual condensate.

Admittedly, it will cost more single particle energy to place particles in
localized wavefunctions like $\phi _{1}$ and $\phi _{2}$ than in $\phi _{0}$%
, so the single particle energy favors the single condensate. Thus we see
that there is a competition between the interaction energy benefit that may
arise from separating the bosons, and the single particle energy cost of
establishing this separation. The role of the trap potential in
fragmentation is now clear: for a non-uniform potential, one can construct
wavefunctions that are confined to the valleys of the potential and which
have little density overlap without a large cost of single particle energy.
A non-uniform potential therefore encourages spatial fragmentation.

\section{Fragmentation in a double well potential}

We now turn to a simple model of the condensate in a potential well that is
symmetric along each Cartesian axis, but which along one axis ($x$) exhibits
a double minimum. We wish to consider which of the single or dual condensate
is energetically favored as a function of the strength of the central
barrier of the potential. As is customary, one minimizes the energy of the
single condensate with respect to variations in $\phi _{0}$ to obtain a
non-linear Schr\"{o}dinger equation (NLSE) for $\phi _{0}.$ Due to the
symmetry of the potential, we expect that the energetically favored dual
condensate will have the particles distributed equally between wavefunctions
that are mirror images of one another across $x=0$, that is, $\phi
_{1}(-x,y,z)=\phi _{2}(x,y,z)$ and $N_{1}=N_{2}=N/2.$ Assuming a dual
condensate of this form, we can minimize its energy with respect to
variations in $\phi _{1}$ and $\phi _{2}$ to obtain two coupled NLSEs for $%
\phi _{1}$ and $\phi _{2}.$

Let us first consider the case of an infinitely strong barrier; all
wavefunctions will then have zero amplitude at the center of the barrier.
The single condensate that minimizes the expectation value of the energy has
a degeneracy associated with the different possible symmetries of the
wavefunction across the barrier. Let us denote the symmetric solution of the
single condensate NLSE as $\phi _{s}$ and the antisymmetric solution as $%
\phi _{a}.$ We can then construct two wavefunctions $\phi _{l}=2^{-1/2}(\phi
_{s}+\phi _{a})$ and $\phi _{r}=2^{-1/2}(\phi _{s}-\phi _{a})$ that are
confined respectively to the left ($x<0$) and right ($x>0$) of the barrier,
and which have the same single particle energy as $\phi _{s}.$

In the single condensate $\left| N,\phi _{s}\right\rangle $, the magnitude
of the interaction energy is $\frac{1}{2}gN(N-1)\int \phi _{s}^{2}({\bf r}%
)d^{3}{\bf r.}$ This can be understood as follows: there are $N$ particles
interacting with $N-1$ others with an interaction strength of $g\int \phi
_{s}^{2}({\bf r})d^{3}{\bf r,}$ and the factor of $1/2$ is due to
over-counting. Now consider the particular dual condensate that is of the
form $\left| N/2,\phi _{l};N/2,\phi _{r}\right\rangle ,$ noting that this
state is {\it not} necessarily the solution of the dual condensate
variational problem. The interaction energy of this state originates solely
from the interaction of particles with others in the same wavefunction,
since the density overlap of the two wavefunctions, $\int \phi _{l}^{2}({\bf %
r})\phi _{r}^{2}({\bf r})d^{3}{\bf r},$ is zero. Particles in this dual
condensate interact twice as intensely with one another than in the single
condensate since $\int \phi _{l}^{4}({\bf r})d^{3}{\bf r}=\int \phi _{r}^{4}(%
{\bf r})d^{3}{\bf r}=2\int \phi _{s}^{4}({\bf r})d^{3}{\bf r.}$ However,
each of the $N$ particles sees only $N/2-1$ others rather than $N-1$ others.
Including the over-counting factor of $1/2$, the total interaction energy of
this dual condensate is $\frac{1}{2}gN(N-2)\int \phi _{s}^{4}({\bf r})d^{3}%
{\bf r.}$ This interaction energy is clearly less than that of the single
condensate. Since the single particle energies are the same for the two
states, it follows that the dual condensate $\left| N/2,\phi _{l};N/2,\phi
_{r}\right\rangle $ is energetically favored overall. Further, the dual
condensate that is obtained from solving the variational problem may have an
energy that is lower still$. $

For small barrier strengths we expect the single condensate to be
energetically favored. This is because a dual condensate that has little
density overlap of $\phi _{1}$ and $\phi _{2}$ will necessarily have a
greater single particle energy than the single condensate. Alternatively, if
the single particle energies are comparable, the wavefunctions $\phi _{1}$
and $\phi _{2}$ will necessarily have a large density overlap. It is
conceivable that if the interaction strength were strong enough, it could
become energetically favorable for the condensate to fragment even in the
absence of any trap potential. However, if the scattering length, $a_{sc}$
is much less than the average width of the ground state of the trap, it can
be verified that a single condensate is preferred at small barrier strengths.

\strut To determine precisely the barrier strength at which the dual
condensate becomes energetically preferred over the single condensate, one
must solve the variational problems for both types of condensate and compare
the energies obtained.

\section{The experimental signature of fragmentation}

The most conspicuous difference between spatially fragmented and
non-fragmented condensates is the extent of spatial coherence, or long range
order, in the system. Specifically, let us imagine a double well potential
with a large central barrier and ask how one could distinguish a single
condensate $\left| N,\phi _{s}\right\rangle $ where $\phi _{s}$ is a doubly
peaked symmetric wavefunction, from a dual condensate $\left| N/2,\phi
_{l};N/2,\phi _{r}\right\rangle $ where $\phi _{l}$ and $\phi _{r}$ are
mirror images of one another centered on the left and right sides of the
barrier respectively. The answer to this question can be found in the recent
literature \cite{Java and Wilkens}; essentially, the difference is revealed
through interference experiments.

A single run of such an experiment proceeds as follows: after the condensate
is formed, the trap potential is removed instantaneously and the particles
are allowed to expand freely until the wavefunctions from either side of the
barrier have overlapped, at which time the positions of the particles are
measured. Assuming the particles do not interact much during the expansion,
the wavefunctions for each particle will evolve independently. If one models
the particle detection in a manner analogous to the standard theory of
photon detection, as is proposed in \cite{Javanainen}, then both the single
and dual condensates will generate an interference pattern. The single
condensate exhibits interference since each particle interferes with itself
just as it would in a double slit experiment. In the dual condensate, the
probability density for each particle is essentially uniform before the
first detection, but as is described in \cite{Javanainen}, the first
detection modifies the state such that the probability densities for the
second and subsequent detections are no longer uniform, and an interference
pattern emerges.

The difference between the single and dual condensates can only be seen by a
comparison of how the spatial phase of the interference pattern varies from
one experimental run to the next. For the single condensate, the spatial
phase is the same in every run, while for the dual condensate, it varies
randomly.

In the experiment of Andrews {\it et al.} \cite{Andrews} a cloud of sodium
atoms was cooled to below the phase-space density threshold for
Bose-Einstein condensation in a double well potential. Upon releasing and
detecting these atoms, an interference pattern was observed. Whether the
spatial phase of interference patterns from different experimental runs are
fixed relative to one another or not has not yet been determined due to
mechanical instabilities in the apparatus. Although it was not our intention
in this paper to provide a detailed model of this experiment, it is clear
that the considerations outlined herein will be relevant to any such model.

\end{document}